\documentclass[a4paper,11pt]{article}
\pdfoutput=1 

\usepackage{jheppub}

\usepackage{color}
\usepackage{graphicx}
\usepackage{amsmath}
\usepackage{amssymb}

\newcommand{\Tr}{\mathrm{Tr}}

\def\Z2{$\mathcal{Z_2}$}

\def\vev#1{\left\langle #1\right\rangle}

\newcommand{\sm}{{Standard Model }}

\def\one{\ensuremath{\mathbf{1}}}
\def\three{\ensuremath{\mathbf{3}} }
\def\threeS{\ensuremath{\mathbf{3^*}} }	

\def\lfv{lepton flavour violation }

\def\TrTrOne{ $\mathrm{ SU(3)_c \otimes SU(3)_L \otimes U(1) }$ }
\def\TrTrTrOne{$\mathrm{SU(3)_c \otimes SU(3)_L \otimes SU(3)_R \otimes U(1)_{X}}$ }

\makeatletter
\def\@fpheader{\relax}
\makeatother

\title{Three-family left-right symmetry with low-scale seesaw mechanism}

\author[a]{Mario Reig}
\author[a]{Jos\'e W.F. Valle}
\author[a]{C.A. Vaquera-Araujo}

\affiliation[a]{AHEP Group, Institut de F\'{i}sica Corpuscular --
  C.S.I.C./Universitat de Val\`{e}ncia, Parc Cient\'ific de Paterna.\\
 C/ Catedr\'atico Jos\'e Beltr\'an, 2 E-46980 Paterna (Valencia) - SPAIN}

\emailAdd{mareiglo@alumni.uv.es}
\emailAdd{valle@ific.uv.es}
\emailAdd{vaquera@ific.uv.es}

\abstract{We suggest a new left-right symmetric model implementing a low-scale
seesaw mechanism in which quantum consistency requires three families of
fermions. The symmetry breaking route to the Standard Model determines
the profile of the ``next'' expected new physics, characterized either
by the simplest left-right gauge symmetry or by the 3-3-1
scenario. The resulting $Z^\prime$ gauge bosons can be probed at the
LHC and provide a production portal for the right-handed neutrinos.
On the other hand, its flavor changing interactions would affect
  the K, D and B neutral meson systems.}

\begin{document}
\maketitle
\flushbottom

\section{Introduction}
\label{motivation}

Both the great success as well as the shortcomings of the Standard
Model (SM) in providing a full picture of reality are well
publicized. Amongst the latter are the facts that it fails in giving a
convincing explanation for small neutrino masses, the existence three
families, and the dynamical origin of parity violation.
Taking these as clues, we suggest a new \sm extension which provides a
dynamical criterion for selecting the profile characterizing the
``next step'' towards physics Beyond the Standard Model.
This would provide an answer to a current burning question, after the
lack of convincing new signals in Run 2 of the LHC: what is the
expected pattern of the upcoming new physics?
By suggesting that these three features are deeply inter-connected we
reconcile the idea of spontaneous parity violation with anomaly
cancellation with, and only with, three families of fermions.

The framework is based on a generalized left-right symmetric
$\mathrm{SU(3)_c} \otimes \mathrm{SU(3)_L} \otimes \mathrm{SU(3)_R}
\otimes \mathrm{U(1)_{X}}$
group structure, first proposed in \cite{Dias:2010vt} in a different
context.
In \cite{Reig:2016tuk} we suggested a unique framework able to mimic
either a usual left-right symmetric theory or a 331 model,
incorporating spontaneous parity violation and small neutrino masses
through the seesaw mechanism.
In order to incorporate the cancellation of the Adler-Bell-Jackiw
anomalies~\cite{Adler:1969gk,Bell:1969ts} the number of fermion
families must chosen to match exactly the number of colors, with
quarks assigned in a non-sequential way, in which one quark family
differs from the other two~\cite{Singer:1980sw}, as in the early
proposal of Refs.~\cite{Singer:1980sw,valle:1983dk} (see
\cite{Fonseca:2016tbn} for an alternative formulation)~\footnote{The
  model is unique up to exotic fermion charge assignments. Unfortunately, the
  interesting choice made in~\cite{Pisano:1991ee,Frampton:1992wt} is
  inconsistent with our construction, due to the perturbativity
  requirement of the gauge couplings \cite{Dias:2010vt}.}.
We choose the simplest scalar sector for which light neutrino masses
are proportional to a small parameter, implementing a genuine
low-scale realization of the seesaw mechanism, such as the
inverse~\cite{Mohapatra:1986bd,GonzalezGarcia:1988rw,Akhmedov:1995ip,Akhmedov:1995vm,Malinsky:2005bi,Catano:2012kw}
or linear seesaw
mechanism~\cite{Akhmedov:1995ip,Akhmedov:1995vm,Malinsky:2005bi},
instead of the more conventional high-scale
seesaw~\cite{GellMann:1980vs,yanagida:1979as,mohapatra:1980ia,Schechter:1980gr,Lazarides:1980nt}. 
Moreover the model suggests that the CKM matrix characterizing quark
weak interactions may have a dynamical origin through a new
``leptophobic'' vacuum expectation value.
The scheme has potentially accessible new neutral gauge bosons
coupling non-diagonally to the quark flavors, inducing flavor changing
neutral currents in the K, D and B neutral meson
systems~\cite{Queiroz:2016gif}. On the other hand such $Z^\prime$'s
can also be searched directly in dilepton studies at the
LHC~\cite{ATLAS-CONF-2015-081}.

This paper is organized as follows. First we sketch our framework
reconciling the existence of parity at a fundamental level with a
gauge structure consistent only for the case of three families of
fermions.
We then show how different patterns of symmetry breaking determine
alternative profiles for the ``next'' expected physics beyond the
Standard Model. Finally we briefly discuss generic features of fermion masses and
mixings, as well as their interactions and phenomenological
implications for the LHC and K, D and B meson mixing.

\section{The model}
\label{sec:model}

The gauge group is \TrTrTrOne with the three families of fermion
fields assigned as,
\begin{equation}
\psi^{a}_{L,R}=\left (
\begin{array}{c}
\nu\\
\ell^{-} \\
N \\
\end{array}\right )^{a}_{L,R}\,,\qquad
Q_{L,R}^\alpha=\left (
\begin{array}{c}
d\\
-u\\
D\\
\end{array}\right )^{\alpha}_{L,R}\,,\qquad
Q_{L,R}^3=\left (
\begin{array}{c}
u\\
d\\
U\\
\end{array}\right )^3_{L,R},
\end{equation}
where parity acts as: $\psi^{a}_{L}\leftrightarrow\psi^{a}_{R}$,
$Q^{\alpha}_L\leftrightarrow Q^{\alpha}_R$,
$Q^{3}_L\leftrightarrow Q^{3}_R$.
\begin{table}[!h]
\begin{center}
\begin{tabular}{|c||c|c||c|c|c|c||c||c|c|c|c|}
\hline
 & $\psi_{L}^{a}$ & $\psi_{R}^{a}$  & $Q_L^{\alpha}$  & $Q_R^{\alpha}$ & $Q_L^3$ & $Q_R^3$  & $S_{L}^{a}$ & $\Phi$ & $\rho$ & $\chi_L$ & $\chi_R$\\
\hline
\hline
$\mathrm{SU(3)_c}$ & \one &\one &\three &\three &\three &\three &\one  &\one&\one &\one &\one \\
\hline
$\mathrm{SU(3)_L}$ & \three & \one & \threeS  & \one & \three & \one &\one & \three  & \three  & \three & \one \\
\hline
$\mathrm{SU(3)_R}$ &\one &\three &\one  &\threeS &\one &\three &\one &\threeS & \three  &\one &\three\\
\hline
$\mathrm{U(1)_{X}}$ & $-\frac{1}{3}$ & $-\frac{1}{3}$  & $0$  & $0$ & $\frac{1}{3}$ & $\frac{1}{3}$& 0 & $0$ & $\frac{1}{3}$ & $-\frac{1}{3}$ & $-\frac{1}{3}$\\ 
\hline
\hline
\end{tabular}
\caption{Particle content of the model, with $a=1,2,3$ and
  $\alpha=1,2$.}
\label{tab:content}
\end{center}
\end{table}

The particle content of the model is summarized in
Table~\ref{tab:content}. We also introduce the fermion fields
$S_{L}^{a}$, transforming as singlets under the
$\mathrm{SU(3)_c} \otimes \mathrm{SU(3)_L} \otimes \mathrm{SU(3)_R}
\otimes \mathrm{U(1)_{X}}$
gauge symmetry and crucial ingredients of low-scale seesaw
schemes~\cite{Mohapatra:1986bd} as well as a set of scalar multiplets
including a bi-triplet scalar $\Phi$ involved in the generation of
charged fermion masses, as well as $\chi_L$ and $\chi_R$ triggering
neutrino masses, see below.
Notice that, following the procedure used in
Refs.~\cite{Akhmedov:1995ip,Akhmedov:1995vm,Malinsky:2005bi}, in the
current model we replace the sextets $\Delta_L$ and $\Delta_R$ present
in the construction of Ref.~\cite{Reig:2016tuk,Huong:2016kpa}, by the fields $\chi_L$
and $\chi_R$
\begin{equation}
  \chi_L \sim(\one, \three , \one , -\frac{1}{3}) ,
  \hspace{3mm}\chi_R \sim(\one, \one , \three , -\frac{1}{3})\,.
\end{equation}
In the presence of the gauge singlet fermions, the Yukawa interactions
for leptons are given by
\begin{equation}
-\mathcal{L}_{\text{leptons}} =\sum_{a,b}\left[y_R^{ab} \overline{\psi}^{a}_R \chi_R S^b_L +y_L^{ab} \overline{\psi}^{a}_L \chi_L (S^{b}_L)^c  +  y^{ab}\overline{\psi}_{L}^{a}\Phi \psi_{R}^{b} +\frac{\mu_{ab}}{2}\overline{(S^a_L)^c} S_L^b \right]+\text{h.c.}
\end{equation}
Parity acts on singlet fermions as $S_L\leftrightarrow(S_L)^c$ and the
scalar fields transform as $\Phi\leftrightarrow \Phi^\dagger$,
$\chi_{L}\leftrightarrow\chi_{R}$, which implies that $y_R=y_L$ and
$y=y^{\dagger}$.  Spontaneous symmetry breaking is driven by the
following structure of vacuum expectation values
(VEVs)~\footnote{We restrict our analysis to the case
    in which there is no mixing at tree level between exotic and SM
    fermions. Note that, for the most general VEV structure a small
    mixing between \sm and exotic fermions is induced, but it will be
    controlled by the ratio $k_{i}/n$ which is expected to be
    small.}:
\begin{equation}\label{eq:pattern}
  \begin{split}
    \vev{\Phi}=\frac{1}{\sqrt{2}}\text{diag}(k_1 ,k_2 ,n)\,,&\qquad
\vev{\rho}=\frac{1}{\sqrt{2}}\left ( \begin{array}{ccc}0&k_3&0\\k_4&0&0\\0&0&0\end{array}\right )\,,\\
  \vev{\chi_R}=\frac{1}{\sqrt{2}}\left ( \begin{array}{c}v_R\\0\\\Lambda_R\end{array}\right )\,, &\qquad
  \vev{\chi_L}=\frac{1}{\sqrt{2}}\left ( \begin{array}{c}v_L\\0\\\Lambda_L\end{array}\right )\,.
    \end{split}
\end{equation}
In this work, we assume for simplicity $k_4, \Lambda_L=0$ and the hierarchy
$n\,,v_R\,,\Lambda_R\,\gg k_{i}\,,v_L$, $i=1,\dots,3$.
After symmetry breaking the electric charge is identified as the unbroken
combination
\begin{equation}\label{elcharge}
Q=T^{3}_{L} +T^{3}_{R}+\beta (T^{8}_{L}+T^{8}_{R}) + X ,
\end{equation}
and the parameter  $\beta$ is given, in our construction, by

\begin{equation}
  \label{eq:beta}
\beta =-1/\sqrt{3}  \,,
\end{equation}
which corresponds to the SVS case~\cite{Singer:1980sw}.
\section{Spontaneous symmetry breaking}
\label{sec:spont-symm-break}

The most general scalar potential characterizing symmetry breaking in
our model can be written as
\begin{eqnarray}
V&=&V_{\chi}+V_\Phi+V_\rho+V_{\Phi\chi}+V_{\Phi\rho}+V_{\chi \rho}+V_{\Phi\rho\chi}\,,\\
V_{\chi}&=& \mu_{\chi}^2( \chi_L^{\dagger}\chi_L+ \chi_R^{\dagger}\chi_R)+\lambda_1 [(\chi_L^{\dagger}\chi_L)^2+ (\chi_R^{\dagger}\chi_R)^2]+\lambda_{2}\chi_L^{\dagger}\chi_L\chi_R^\dagger\chi_R\hspace{1mm},\nonumber\\
V_\Phi&=&\mu_\Phi^2 \Tr(\Phi\Phi^{\dagger})+f_\Phi[\epsilon^{l_1 l_2 l_3}\epsilon_{r_1 r_2 r_3}\Phi_{l_1}^{r_1}\Phi_{l_2}^{r_2}\Phi_{l_3}^{r_3}+ \mathrm{h.c.}]+\lambda_3 \Tr(\Phi\Phi^{\dagger})^2+\lambda_4  \Tr(\Phi\Phi^{\dagger}\Phi\Phi^{\dagger})\hspace{1mm},\nonumber\\
V_\rho&=&\mu_\rho^2 \Tr(\rho\rho^{\dagger})+\lambda_5 \Tr(\rho\rho^{\dagger})^2+\lambda_6  \Tr(\rho\rho^{\dagger}\rho\rho^{\dagger})\hspace{1mm},\nonumber\\
V_{\Phi\chi}&=&
f_{\chi} (\chi^\dagger_L\Phi\chi_R+\mathrm{h.c.}) +\lambda_{7}(\chi_L^\dagger\chi_L+\chi_R^\dagger\chi_R)\Tr(\Phi\Phi^\dagger)+\lambda_{8}\big (\chi_L^\dagger\Phi\Phi^\dagger\chi_L+\chi_R^\dagger\Phi^\dagger\Phi\chi_R \big ) \,,\nonumber\\
V_{\Phi\rho}&=&\lambda_{9}\Tr(\rho\rho^\dagger )\Tr(\Phi \Phi^\dagger)+\lambda_{10}\big [ \Tr(\Phi\Phi^\dagger\rho\rho^\dagger )+\Tr (\Phi^\dagger\Phi\rho^T\rho^*)\big ]\nonumber\\&&+\lambda_{11}\big [ \Tr(\Phi^\dagger\rho\Phi^T \rho^* )+\Tr(\Phi\rho^T\Phi^* \rho^\dagger )   \big ]\,,\nonumber\\
V_{\chi\rho}&=& \lambda_{12}\Tr(\rho\rho^\dagger)(\chi^\dagger_L\chi_L+\chi^\dagger_R\chi_R)    +\lambda_{13}(\chi^\dagger_L\rho\rho^\dagger\chi_L+\chi^\dagger_R\rho^T\rho^*\chi_R)   \,,\nonumber\\
V_{\Phi\rho\chi}&=&f_{\rho}[\epsilon^{l_1 l_2 l_3}(\Phi\rho^T)_{l_1 l_2}(\chi_L)_{l_3}+\epsilon_{r_1 r_2 r_3}(\Phi^\dagger\rho)^{r_1 r_2}(\chi_R)^{r_3}+\mathrm{h.c.}]\nonumber\\&&+\lambda_{14}(\Phi\rho^T\Phi\chi_R+\Phi^\dagger\rho\Phi^\dagger\chi_L+\mathrm{h.c.})\,,\nonumber
\end{eqnarray}

Minimization leads to the following tadpole equations:
\begin{equation}\label{tad1}\begin{split}
\mu _{\Phi }^2=&- \left(n^2+k_1^2+k_2^2\right)\lambda_3- (n^2+k_2^2)\lambda_4-\frac{1}{2} 
   \bigg [k_3^2\lambda_9-\frac{k_2^2k_3^2}{n^2-k_2^2}\lambda_{10}+\frac{n^2\Lambda_R^2}{n^2-k_2^2}\lambda_8+(v_L^2+v_R^2+\Lambda_R^2)\lambda_7\bigg]\,,\\
   \mu _{ \rho}^2=&-\frac{1}{2}\bigg [\frac{v_L^2\Lambda_R^2(v_L^2-v_R^2-\Lambda_R^2)}{k_3^2(v_L^2-v_R^2)}(2\lambda_1-\lambda_2)+ k_3^2(\lambda_5+\lambda_6)+\frac{(n^2-k_1^2)\Lambda_R^2}{k_3^2}\lambda_8+(k_1^2+k_2^2+n^2)\lambda_9\\&+(k_1^2+k_2^2)\lambda_{10}+(v_L^2+v_R^2+\Lambda_R^2)\lambda_{12}+v_L^2\left(1+\frac{\Lambda_R^2}{v_L^2-v_R^2}\right)\lambda_{13}\bigg ]\,,\\
\mu _{ \chi}^2=&- \left(v_L^2+v_R^2+\frac{v_R^2\Lambda_R^2}{v_R^2-v_L^2}\right)\lambda_1-\frac{1}{2}\bigg [\frac{\Lambda_R^2v_L^2}{(v_L^2-v_R^2)}\lambda_{2}+(k_1^2+k_2^2+n^2)\lambda_7+ 
   k_1^2\lambda_8+k_3^2\lambda_{12}\\&+\frac{k_3^2v_L^2}{(v_L^2-v_R^2)}\lambda_{13}\bigg ]\,,\\
f_\Phi=&\frac{2(n^2-k_2^2)nk_2\lambda_4-nk_2k_3^2\lambda_{10}+k_3\Lambda_R(n^2-k_2^2)\lambda_{14}+nk_2\Lambda_R^2\lambda_8}{\sqrt{2}k_1(v_L^2-v_R^2)}\,,\\
f_{\rho}=& \frac{1}{\sqrt{2}k_1}\bigg \{ nk_2\lambda_{14}+\frac{\Lambda_R(k_1^2-n^2)}{k_3}\lambda_8+\frac{v_L^2\Lambda_R}{k_3(v_L^2-v_R^2)}[k_3^2\lambda_{13}+(v_L^2-v_R^2-\Lambda_R^2)(2\lambda_1-\lambda_2)]  \bigg \}   \,,\\
f_{\chi}=&\frac{v_Lv_R\big [ (2\lambda_1-\lambda_2)(v_R^2+\Lambda_R^2-v_L^2) -k_3^2\lambda_{13} \big ]}{\sqrt{2}(v_R^2-v_L^2)k_1}\,,
\end{split}
\end{equation}
together with the relations
\begin{equation}\label{tad2}\begin{split}
&
2k_1^2k_2k_3^2(v_L^2-v_R^2)\lambda_{11}-nv_Lv_R\Lambda_R^2[k_3^2\lambda_{13}+(2\lambda_1-\lambda_2)(v_L^2-v_R^2-\Lambda_R^2)]=0\,\\
&
nk_3(k_1^2-k_2^2)\lambda_{14}+(n^2-k_1^2)k_2\Lambda_R\lambda_8+\frac{k_3^2v_L(nv_R-k_2v_L)\Lambda_R}{v_L^2-v_R^2}\lambda_{13}\\&\qquad+\frac{v_L(k_2v_L-nv_R)\Lambda_R(v_L^2-v_R^2-\Lambda_R^2)}{v_L^2-v_R^2}(\lambda_2-2\lambda_1)=0\,,\\
&
(k_1^2-k_2^2)\left[ (k_1^2-n^2)\lambda_4+\frac{n^2k_3^2}{(n^2-k_2^2)}\lambda_{10} \right]+\frac{\lambda_8}{2}\bigg \{ k_1^2(v_L^2+v_R^2)-\frac{[n^4+(k_1^2-2n^2)k_2^2]\Lambda_R^2}{n^2-k_2^2} \bigg \}\\&\qquad+nk_2k_3\Lambda_R\lambda_{14}+\frac{v_R^2+\Lambda_R^2}{2(v_L^2-v_R^2)}\big [ k_3^2v_L^2\lambda_{13}+v_L^2(v_L^2-v_R^2-\Lambda_R^2)(2\lambda_1-\lambda_2)\big ]=0  \,.
\end{split}
\end{equation}
Different patterns of symmetry breaking associated to
different hierarchies of the above vacuum expectation values will thus
translate into a different nature for the physics beyond the \sm
expected above current LHC energies. In one of them the ``next''
theory will be left-right symmetric, while in the other it will follow
the pattern dictated by the 331 gauge dynamics.

\begin{figure}
  \centering
    \includegraphics[width=0.7\textwidth]{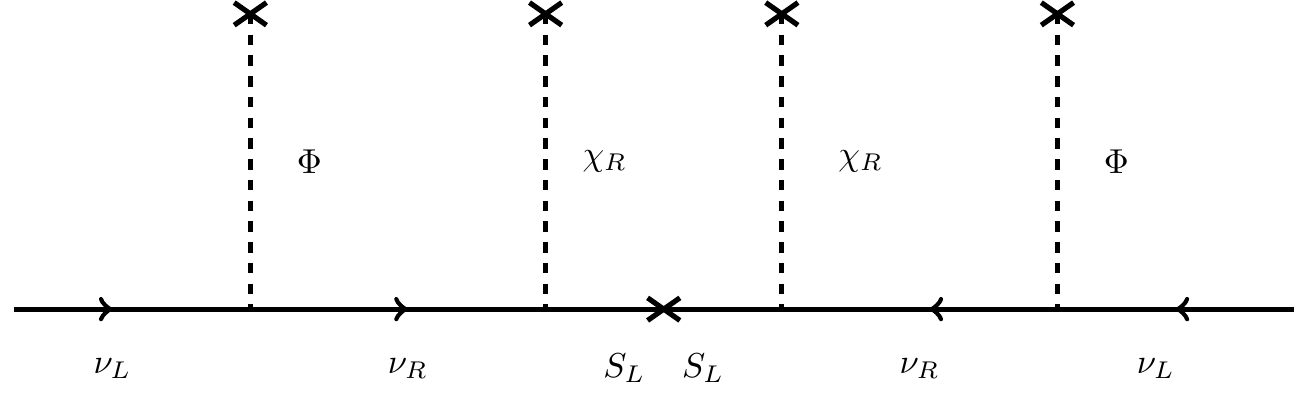}
    \caption{Inverse seesaw diagram. }
  \label{InvSS}
\end{figure}

\begin{figure}
  \centering
    \includegraphics[width=0.5\textwidth]{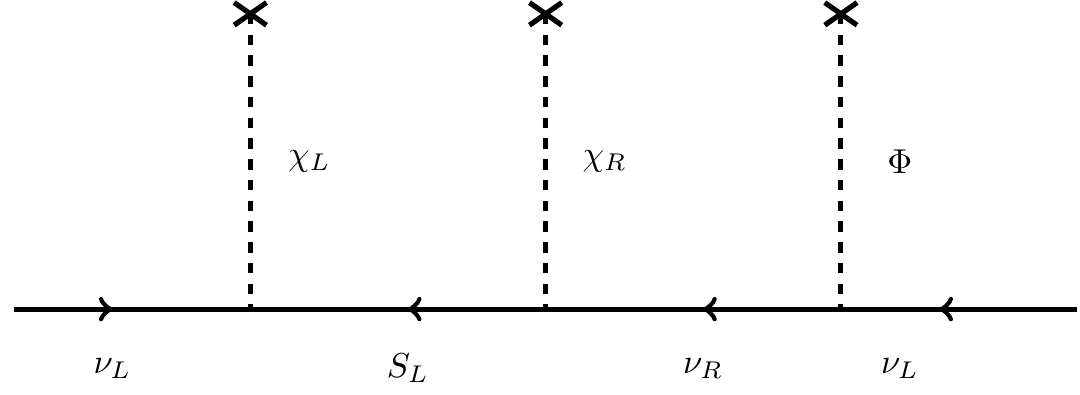}
    \caption{Linear seesaw diagram, up to symmetrization. }
  \label{LinSS}
\end{figure}

\section{Phenomenological implications}
\label{sec:phenomenology}

After spontaneous symmetry breaking driven by the pattern in
Eq.~(\ref{eq:pattern}) the neutrino mass matrix in the
$(\nu_L,\nu_R^c, N_L, N_R^c, S_L)$ basis is given by:
\begin{equation}
M_\nu=\left(\begin{array}{ccccc}
0  & y k_1  & 0 & 0  & y_L v_L  \\
y^T k_1  & 0 & 0 & 0  & y_R v_R \\
0 & 0 & 0 & y\, n  & 0  \\
0 & 0 & y^T n  &  0  & y_R \Lambda_R \\
y_L^T v_L & y_R^T v_R & 0  &  y_R^T \Lambda_R  &   \mu \\
\end{array}\right).
\end{equation}
For example, setting $v_L\to 0$ gives us 
\begin{equation}
m_{light}\approx\frac{k_1^2}{v_R^2} (y\,y_R^{T-1})\mu (y_R^{-1}\,y^T)\,,
\end{equation}
which vanishes as $\mu \to 0$, reproducing the standard inverse seesaw
relation~\cite{Mohapatra:1986bd,GonzalezGarcia:1988rw} illustrated in
Fig.~\ref{InvSS}.
On the other hand, taking $\mu \to 0$ leads to
\begin{equation}
m_{light}\approx\frac{v_L k_1}{v_R} \left[y(y_R\,y_L^{-1})^T+ (y_R\,y_L^{-1}) y^T\right]\,,
\end{equation}
recovering the linear seesaw mass relation~\cite{Malinsky:2005bi}, see
Fig.~\ref{LinSS}.  Besides, as dictated by parity, the Yukawa
couplings satisfy $y_L=y_R$ and the above mass relation reduces to
$m_{light}\approx\frac{v_L k_1}{v_R} (y+ y^T)$. In either case we can
naturally explain the lightness of neutrinos as a result of the
exchange of quasi-Dirac heavy lepton messengers that could lie in the
TeV scale~\cite{Dittmar:1989yg,Deppisch:2013cya}. The required mixing
parameters measured in neutrino oscillation experiments can also be
accommodated.

In the quark sector, the up- and down-type quark mass matrices are
determined by the Yukawa interactions
\begin{equation}
\label{lagyukq}
\begin{split}
-\mathcal{L}_{\text{quarks}}=&\sum _{\alpha,\beta=1}^{2}\left(h^{Q}_{\alpha\beta}\overline{Q}_{L}^{\,\alpha}\Phi^* Q_{R}^{\beta}\right)+h^{Q}_{33}\overline{Q}^{\,3}_{L}\Phi Q^3_{R}+\sum _{\alpha=1}^{2}\left(h^{Q}_{\alpha 3}\overline{Q}_{L}^{\,\alpha}\rho^* Q_{R}^{3}+h^{Q}_{3\alpha}\overline{Q}^{\,3}_{L}\rho Q^\alpha_{R}\right)+\mathrm{h.c.}
\end{split}
\end{equation}
and are given by
\begin{equation}\label{mass_up}
\begin{split}
M^{u}&=\frac{1}{\sqrt{2}}\left(
\begin{array}{ccc}
h_{11}^Qk_2 &  h_{12}^Qk_2 &  0  \\
 h_{21}^Qk_2 &  h_{22}^Qk_2 &  0 \\
-h_{31}^Q k_3 &  -h_{32}^Q k_3 &  h_{33}^Qk_1 \\
\end{array}
\right)\,,
\end{split}
\end{equation}
\begin{equation}\label{mass_down}
\begin{split}
M^{d}&=\frac{1}{\sqrt{2}}\left(
\begin{array}{ccc}
 h_{11}^Qk_1 &  h_{12}^Qk_1 & h_{13}^Q k_3 \\
 h_{21}^Qk_1 &  h_{22}^Qk_1 & h_{23}^Q k_3 \\
0 & 0 &  h_{33}^Qk_2 \\
\end{array}
\right)\,.
\end{split}
\end{equation}

Concerning the neutral current weak interaction, these are exactly
flavor-diagonal as expected by the Glashow-Iliopoulos-Maiani
mechanism.
There are, however, new neutral gauge bosons that couple
non-diagonally to the quark flavors. 
Such flavor changing interactions are constrained by direct searches
at the LHC Dilepton data, as well as by limits from K, D and B neutral
meson systems~\cite{Queiroz:2016gif}.
On the other hand right-handed neutrino pair production through the
$Z^\prime$ gauge boson portal would induce sizeable rates for \lfv
effects at the high energies available at the
LHC~\cite{Deppisch:2013cya}, even when processes like
$\mu \to e \gamma$ are negligibly small.

\section{Flavor-changing neutral currents}

An interesting feature of our model is the break down of the
Glashow-Iliopoulos-Maiani mechanism in the weak neutral current.
Indeed there are flavor-changing neutral currents (FCNC) mediated by
the new heavy neutral gauge bosons $Z^\prime_L$ and $Z^{\prime}_R$
associated to the left and right SU(3) groups. The neutral current
Lagrangian is given by
\begin{equation}\begin{split}
\mathcal{L}_{NC}=&-\overline{f}_{Li}\gamma^\mu\left[g_L(T_3^LW^{L3}_\mu+T_8^LW^{L8}_\mu)+g_xX_iB_\mu\right]f_{Li}\\&
-\overline{f}_{Ri}\gamma^\mu\left[g_L(T_3^RW^{R3}_\mu+T_8^RW^{R8}_\mu)+g_xX_iB_\mu\right]f_{Ri}\,,
\end{split}\end{equation}
with
  $X=Q-T^3_L-T^3_R+\frac{1}{\sqrt{3}}(T^8_L+T^8_R)$. We note that the
  \sm neutral gauge bosons, the photon and the standard $Z$ boson, do
  not mediate FCNC since their coupling involves only the electric
  charge and $T^3$ generators in flavor space, hence proportional to
  the identity matrix. However the $T_8$ piece of the neutral current
  leads to flavor change, and taking this into account we obtain the
  FCNC piece of the neutral current Lagrangian,
\begin{equation}\begin{split}
\mathcal{L}_{\text{FCNC}}\supset &-\overline{f}_{Li}\gamma^\mu T_8^L(g_L W_\mu^{L8}+\frac{1}{\sqrt{3}} g_XB_\mu)f_{Li}\\&
-\overline{f}_{Ri}\gamma^\mu T_8^R(g_L W_\mu^{R8}+\frac{1}{\sqrt{3}} g_XB_\mu)f_{Ri}  \,.
\end{split}\end{equation}
Defining the relevant fields $Z^\prime_{L,R \mu}$ and couplings $g_{L,R}^\prime$ as
\begin{equation}\begin{split}
&g_L^\prime Z^\prime_{L\mu}=g_LW_\mu^{L8}+\frac{1}{\sqrt{3}} g_XB_\mu\,,\\&
g_R^\prime Z^\prime_{R\mu}=g_LW_\mu^{R8}+\frac{1}{\sqrt{3}} g_XB_\mu\,,
\end{split}
\end{equation}
we can write down the FCNC Lagrangian 
\begin{equation}
\begin{split}
\mathcal{L}_{\text{FCNC}}
&=-\frac{1}{\sqrt{3}}\left[g^\prime_L Z^\prime_{L\mu}(V^{qL}_{3j})^* V^{qL}_{3i}\overline{q_L^{\prime j}}\gamma^\mu q_L^{\prime i}+g^\prime_R Z^\prime_{R\mu}(V^{qR}_{3j})^* V^{qR}_{3i}\overline{q_R^{\prime j}}\gamma^\mu q_R^{\prime i}\right]\,,
\end{split}
\end{equation}
with $q'=(u,c,t)$ \textbf{or $q'=(d,s,b)$ and $i\neq j$}. Thus, the
effective Lagrangian for meson mixing becomes
\begin{equation}
\mathcal{L}^{\text{eff}}_{\text{FCNC}}=\frac{1}{3}\left|(V^{qL}_{3j})^*V^{qL}_{3i}\right|^2\Big ( \frac{g^{\prime 2}_L}{m_{Z_L^\prime}^2}\Big )(\overline{q_L^{\prime j}}\gamma^\mu q_L^{\prime i})^2+\frac{1}{3}\left|(V^{qR}_{3j})^*V^{qR}_{3i}\right|^2\Big ( \frac{g^{\prime 2}_R}{m_{Z_R^\prime}^2}\Big )(\overline{q_R^{\prime j}}\gamma^\mu q_R^{\prime i})^2\,.
\end{equation}
The phenomenology of the model is quite rich since the $Z_{L,R}^\prime$
mediate meson mixing and can also be searched directly at the LHC by
Drell Yann production in the dilepton channel.
One notices that the relative importance of the left and right terms
is determined by the hierarchy of vacuum expectation values which
characterizes the physics ``just beyond'' the \sm in our scheme.
Indeed, these gauge boson mass parameters scale as
$m_{Z_L^\prime}\sim n$ and $m_{Z_R^\prime}\sim \sqrt{n^2+v_R^2}$ so
that for $n \gg v_R$ the intermediate scale physics is described by
the $\mathrm{SU(2)_L \otimes SU(2)_R \otimes U(1)}$ gauge
structure~\footnote{In contrast to the most popular left-right
  symmetric model here the low-scale is consistent with neutrino mass
  generation. }.  In this left-right symmetric limit the FCNC will be
suppressed by the high value of $v_R$.
In contrast, for $n \ll v_R$ one has intermediate\TrTrOne physics at
``low'' scale. In this case one recovers the situation considered
in~\cite{Queiroz:2016gif} with FCNC suppressed by the magnitude of the
vacuum expectation value $n$.
For the generic case of potentially comparable magnitudes for $n$ and
$v_R$ one can generalize the analysis given in~\cite{Queiroz:2016gif}
so as to constrain them jointly, by combining restrictions from meson
mixing with LHC bounds from direct search. These will restrict the
symmetry breaking path from our original ``mother'' theory.

Before concluding this brief sketch, let us mention that, since all
leptons are in the \threeS representation, there are no tree level
flavor changing effects mediated by $Z^\prime_L$ nor $Z^{\prime}_R$ 
in this sector and, as a result, \lfv is adequately suppressed.
The same occurs for exotic (third component) quarks: since the up-type
belong to the \three and down-type transform as \threeS they do not
give rise to flavor change. 

\section{Conclusion}
\label{sec:conclusion}

In summary, we have proposed a framework where the origin of small
neutrino mass, spontaneous parity violation and the existence of three
fermion families all have a common origin, with neutrinos getting mass
from low-scale seesaw, such as the inverse or the linear seesaw
mechanism.
The framework specifies dynamically whether the nature of the new
physics to show up in the next run of the LHC will follow a left-right
symmetric pattern, or a chiral
$\mathrm{SU(3)_c \otimes SU(3)_L \otimes U(1)}$ symmetry profile, with
quark flavor changing interactions associated to new $Z^\prime$ gauge
bosons.
These patterns follow from minimizing the Higgs potential for
different hierarchies of the relevant vacuum expectation values.
The theory reconciles the existence of parity at a fundamental level
with the quantum consistency requirement of having precisely three
families of fermions, in which quarks must be assigned in a
non-sequential way with the third family different from the others.
The existence of a $Z^\prime$ gauge boson would be probed in dilepton
studies, and also provide a production portal for the right-handed
neutrinos responsible for generating neutrino mass via a genuine
low-scale seesaw mechanism.
These new $Z^\prime$ gauge bosons could also induce sub-weak flavor
changing transitions in the K, D and B neutral meson systems.

\section{Acknowledgements}

Work supported by Spanish grants FPA2014-58183-P, Multidark
CSD2009-00064, SEV-2014-0398 (MINECO) and PROMETEOII/2014/084
(Generalitat Valenciana).  C.A.V-A. acknowledges support from CONACyT
(Mexico), grant 274397.  M.~R. was supported by JAEINT-16-00831.

\bibliographystyle{JHEP}
\bibliography{3331,merged_Valle,newrefs}

\end{document}